\documentclass[final,3p,times]{elsarticle}

\usepackage{amssymb}

\usepackage{amssymb}
\usepackage[]{amsmath}
\usepackage{graphics}
\usepackage{amssymb}
\usepackage[]{amsmath}
\usepackage{amsthm}
\usepackage{setspace}
\usepackage{epsfig}
\usepackage{subfigure}
\usepackage{booktabs}
\usepackage{float}
\usepackage{graphicx}
\makeatletter

\newcommand{\Rmnum}[1]{\expandafter\@slowromancap\romannumeral #1@}
\makeatother

\biboptions{numbers,sort&compress}
\usepackage[dvipdfm,colorlinks,linkcolor=red,anchorcolor=blue,citecolor=green]{hyperref} %

\usepackage{amsmath}
\usepackage{times}
\usepackage{anysize}
\marginsize{2.5cm}{2.5cm}{1cm}{1cm}

\selectfont   

\journal {}

\begin{document}

\begin{frontmatter}



\title{Darboux transformation of nonisospectral coupled Gross-Pitaevskii equation and its
multi-component generalization }

%
%

\author[]{Tao Xu$^{1, 2}$}
\author[]{Yong Chen$^{1, 2}$\corref{mycorrespondingauthor}}
\cortext[mycorrespondingauthor]{Corresponding author}
\ead{ychen@sei.ecnu.edu.cn}
\address{1 Shanghai Key Laboratory of Trustworthy Computing, East China Normal University, Shanghai, 200062, China\\
2 MOE International Joint Lab of Trustworthy Software, East China Normal University, Shanghai, 200062, China}

\begin{abstract}
We extend one component Gross-Pitaevskii equation to two component coupled case with the damping term, linear and parabolic density profiles,
then give the Lax pair and infinitely-many conservations laws of this coupled system. The system is nonautonomous, that is, it admits a
nonisospectral linear eigenvalue problem. In fact, the Darboux transformation for this kind of inhomogeneous
system which is essentially different from the isospectral case, we reconstruct the Darboux transformation for this coupled Gross-Pitaevskii equation. Multi nonautonomous solitons, one breather and the first-order rogue wave are also obtained by
the Darboux transformation. When $\beta >0$, the amplitudes and velocities of solitons decay exponentially as $t$ increases, otherwise, they increase exponentially as $t$ increases. Meanwhile, the real part $Re(\xi_j)$'s~$(j=1,2,3,\dots)$ of new spectral parameters determine the direction of solitions' propagation and $\alpha$ affects the localization of solitons. Choosing $Re(\xi_1)=Re(\xi_2)$, the two-soliton bound state is obtained. From nonzero background seed solutions, we construct one nonautonomous breather on curved background and find that this breather has some deformations along the direction of $t$ due to the exponential decaying term. Besides, $\beta$ determines the degree of this curved background, if we set $\beta>0$, the amplitude of the breather becomes small till being zero as $t$ increases. Through taking appropriate limit about the breather, the first-order rogue wave can be acquired. Finally, we give multi-component
generalization of Gross-Pitaevskii equation and its Lax pair with nonisospectral parameter, meanwhile, Darboux transformation about this
multi-component generalization is also constructed.
\end{abstract}


\end{frontmatter}



\section{Introduction}
In recent years, analytic solutions of the nonlinear evolution equations, such as solitons \cite{3-1,3-2,3-3,3-4}, breathers \cite{3-5,3-6,3-7} and rogue waves \cite{3-8,3-9,3-10} and so on, which have received a large of research activities in many realms. When the effect of dispersion and nonlinearity is balanced in nonlinear waves during propagating, solitons will be formed. These waves which keep their features (amplitudes, speeds and so on) unchanged, during they propagating and after interacting with each other. In many cases, they are considered as the ideal solution models in physics \cite{3-11}. As the particular solutions of nonlinear systems, breathers propagate steadily and localize in either time or space, especially Akhmediev breather (AB) \cite{3-12,3-13} and Kuznetsov-Ma breather (KM) \cite{3-14}. Another special type of analytic solution are the rogue waves which localized in both space and time, and they have peak amplitudes usually more than twice the background wave height \cite{3-15,3-16}. Besides, rogue waves appear from nowhere and disappear without a trace, through mathematical methods, they can be also written in terms of the rational functions of coordinates.

In many documents, these above three types of analytical solutions have been discovered in the autonomous systems, whose coefficients are all constants \cite{3-17,3-18}. However, for the practical situations, there are many models with variable-coefficients, in other words, they conclude time or (both) space-dependent nonlinearity, dispersion and external potentials. So, these systems are nonautonomous \cite{3-19,3-20,3-21}. Properties of these analytic solutions in the nonautonomous systems are greatly different from those in autonomous ones. For example, when these solitons propagate, their amplitudes, speeds and widths are all changed \cite{3-22,3-23}.

In the plasma, when both high frequency and low frequency waves that have the same group velocity are excited, they can be nonlinearly coupled by density depletion and through the generation of high frequency side bands and low frequency second harmonic. So, in this case, the waves are suggested to be determined by the coupled inhomogeneous Gross-Pitaevskii (G-P) equation \cite{3-19,3-24,3-25}, the system can be written as
\begin{equation}\label{xt-3-1}
\begin{array}{l}
iq_{1t}+q_{1xx}+2\mu^2(|q_1|^2+|q_2|^2)q_1+(i\beta-\alpha x+\beta^2x^2)q_1=0,\\
iq_{2t}+q_{2xx}+2\mu^2(|q_1|^2+|q_2|^2)q_2+(i\beta-\alpha x+\beta^2x^2)q_2=0,
\end{array}
\end{equation}
where $q_1(x,t)$ and $q_2(x,t)$  are the complex envelops of two fields in the inhomogeneous plasma, $\mu$,~$\alpha$,~$\beta$ are all real numbers, $\mu$ is the nonlinearity parameter, $\alpha x$ stands for linear density profile and $\beta^2x^2$ corresponds to parabolic density profile, meanwhile, $i\beta$ is the damping term. Setting $\mu=1$ and $\alpha=\epsilon$, Eq.(1) becomes the coupled systme in \cite{3-25}, where the authors gave soliton solutions by B\"{a}cklund transformation and its $\emph{N}$-coupled damped generalization.

 In \cite{3-19}, one-component G-P is studied by Darboux transformation (DT) and its multi solitons, breather and higher rogue waves are obtained, but, we verify its DT by Maple software and find that this kind of DT does not admit the t-part of Lax pair. So, we reconstruct the DT of coupled G-P equation in this paper. In \cite{3-26}, the authors verified that the Darboux transformation about a generalized inhomogeneous higher-order nonlinear Schr\"{o}dinger equation was incorrect in \cite{3-27}. Because this inhomogeneous equation which also admitted a nonisospectral nonlinear eigenvalue problem was mistaken for having a constant spectral parameter by the authors in \cite{3-27}. Modified DT about the same inhomogeneous higher-order nonlinear Schr\"{o}dinger equation is reconstructed and some novel solitons are also given in \cite{3-26}. There have been many other articles about one- and multi-component nonautonomous systems. The variable-coefficients coupled Hirota equation was studied in \cite{3-28}, then some new types
rogue waves were obtained by the authors, such as dark-bright, composite, three-sister, quadruple and sextuple rogue waves. In \cite{3-24}, some bright solitons of the system in a nonlinear inhomogeneous fiber were obtained by Hirota bilinear method through constructing double wronskian determinant\cite{3-29}. Utilizing the same method, nonautonomous matter waves were also constructed in a spin-1 Bose-Einstein condensate\cite{3-30}.

In this paper, we extend one-component G-P equation\cite{3-19} to two-component case, and reconstruct the DT of this coupled system with nonisospectral linear eigenvalue problem. Eq.(\ref{xt-3-1}) admits the AKNS (Ablowitz-Kaup-Newell-Segur) hierarchy and Gu has constructed a unified approach \cite{3-31} to construct Darboux transformation for this isospectral AKNS, however, Eq.(\ref{xt-3-1})'s spectral parameter is  nonisospectral. So, we can not directly utilize Gu's method. Based on Gu's method, Zhou \cite{3-32,3-33} generalized Gu's formula to the nonisospectral cases. In the following contents, based on Zhou's method, we reconstruct Darboux transformation about this coupled system (\ref{xt-3-1}) and its multi-component generalization \cite{3-25}. The spectral parameter $\lambda$ of Eq.(\ref{xt-3-1}) holds $\lambda=\frac{\alpha}{4\beta}+\xi e^{-2\beta t}$  with $\xi$ being an arbitrary constant, thus we can take it as a new spectral  parameter \cite{3-19}, in order to make sure its integrability, we present  Eq.(\ref{xt-3-1})'s infinitely-many conservation laws. Utilizing the Darboux transformation which is constructed by us, the multi nonautonomous solitons, one breather and the first-order rogue wave are constructed. For multi nonautonomous solitons, several free parameters play some important roles in soliton solutions structures, namely, $\mu$  and the imaginary part of spectral parameters $Im(\xi_j)$'s~$(j=1,2,3,\dots)$ determine the amplitudes of solitons, then, the real part of spectral parameters $Re(\xi_j)$'s affect the direction of solitions' propagation
and $\alpha$ impacts the localization of solitons. When choosing $Re(\xi_1)=Re(\xi_2)$, the two-soliton bound state is obtained. Furthermore, starting from nonzero background seed solutions, nonautonomous breather which based on curved background is acquired, and we can find that this breather has some deformations along the direction of $t$ due to the exponential decaying term. For convenience, we set~$\beta>0$, the parameter $\beta$ determines the degree of this curved background. The amplitude of the breather decreases as $t$ increases till being zero. Through taking appropriate limit in the expressions of breather, we also get the first-order rogue wave based on curved background. Meanwhile, the multi-component generalization of Gross-Pitaevskii equation and its Lax pair are all also given, then we construct its DT based on the method which used in two-component case.

The paper is organized as follows. In section 2, the Lax pair and conservation laws of Eq.(\ref{xt-3-1}) are given.
In section 3, the Darboux transformation of Eq.(\ref{xt-3-1}) with nonisospectral Lax pair is constructed and multi solitons are also obtained. In section 4, breather and rogue waves are revealed. In section 5, the Darboux transformation of multi-component generalization of Eq.(\ref{xt-3-1}) is discussed. The last section contains several conclusions and discussions.

\section{Integrability: Lax pair and conservation laws}
Based on \cite{3-19} and \cite{3-25}, the Lax pair of Eq.(\ref{xt-3-1}) can be written as follow :
\begin{eqnarray}
&&\Psi_{x}=U\Psi=(i\lambda U_0+\mu U_1)\Psi,\label{xt-3-2}\\
&&\Psi_{t}=V\Psi=(2i\lambda^2 U_0+2\lambda( -i\beta x U_0+\mu U_1)+iV_1)\Psi,\label{xt-3-3}
\end{eqnarray}
Where
\begin{gather*}
U_0=\begin{bmatrix}-1&0&0\\0&1&0\\0&0&1\end{bmatrix},\quad
U_1=\begin{bmatrix}0&Q_1&Q_2\\-Q_1^*&0&0\\-Q_2^*&0&0 \end{bmatrix},
\end{gather*}

\begin{gather*}
V_1=\begin{bmatrix}\mu^2(|Q_1|^2+|Q_2|^2)-\dfrac{\alpha x}{2}&\mu Q_{1x}+2i\mu \beta xQ_1&\mu Q_{2x}+2i\mu \beta xQ_2\\
                    \mu Q_{1x}^*-2\mu i\beta xQ_1^* &-\mu^2 |Q_1|^2 +\dfrac{\alpha x}{2} &-\mu^2 Q_2Q_1^*\\
                    \mu Q_{2x}^*-2\mu i\beta xQ_2^* &-\mu^2 Q_1Q_2^* &-\mu^2|Q_2|^2 +\dfrac{\alpha x}{2} \end{bmatrix},
\end{gather*}
with $Q_1(x,t)=q_1(x,t)e^{-\frac{i\beta x^2}{2}}$, $Q_2(x,t)=q_2(x,t)e^{-\frac{i\beta x^2}{2}}$ and $\lambda(t)=\frac{\alpha}{4 \beta}+\xi e^{-2 \beta t}$~($\xi$ is an arbitrary complex constant). Here, $*$ denotes the complex conjugate, and $\lambda$ is dependent on time, so the Lax pair is nonisospectral case. The column vector $\Psi=(\psi_1,\psi_2,\psi_3)^T$ is the eigenfunction with $\lambda$ being the spectral parameter. Besides, the coupled system Eq.(\ref{xt-3-1}) can be directly figured out by the compatibility condition $U_t-V_x+[U,V]=0$.

In the following, we give the infinitely-many conservation laws of Eq.(\ref{xt-3-1}), which further indicates this system is integrable\cite{3-34,3-35,3-36}. For this purpose, we introduce these two functions $\Gamma_1(x,t)=\dfrac{\psi_2}{\psi_1}$ and $\Gamma_2(x,t)=\dfrac{\psi_3}{\psi_1}$, according to Eq. (\ref{xt-3-2}), the following two equalities can be derived, namely
\begin{eqnarray}
&&\Gamma_{1x}=-\mu Q_1^*+2i\lambda \Gamma_1-\mu Q_1 \Gamma_1^2-\mu Q_2\Gamma_1\Gamma_2,\label{xt-3-4}\\
&&\Gamma_{2x}=-\mu Q_2^*+2i\lambda\Gamma_2-\mu Q_2\Gamma_2^2-\mu Q_1\Gamma_1\Gamma_2.\label{xt-3-5}
\end{eqnarray}

Setting $Q_1\Gamma_1=\sum^{\infty}_{n=1}\Gamma_1^{(n)}\lambda^{-n}$ and $Q_2\Gamma_2=\sum^{\infty}_{n=1}\Gamma_2^{(n)}\lambda^{-n}$, where~ $\Gamma_1^{(n)}$'s and $\Gamma_2^{(n)}$'s ~$(n=1,2,3,\dots)$  are all functions of $x$ and $t$ to be determined, and we substitute these two expansions into Eq.(\ref{xt-3-4}) and Eq.(\ref{xt-3-5}) respectively, then equate the coefficients of the same power of $\lambda$, get
 \begin{subequations}\label{xt-3-6}
 \begin{align}
 \lambda^0:&\Gamma_1^{(1)}=-\frac{i}{2}\mu Q_1Q_1^*,\\
 \lambda^{-1}:&\Gamma_1^{(2)}=-\frac{i}{2}(\Gamma_{1x}^{(1)}-\frac{Q_{1x}}{Q_1}\Gamma_1^{(1)})=-\frac{1}{4}\mu Q_1Q_{1x}^*,\\
 \lambda^{-2}:&\nonumber \Gamma_1^{(3)}=-\frac{i}{2}[-\frac{Q_{1x}}{Q_1}\Gamma_1^{(2)}+\Gamma_{1x}^{(2)}+\mu\Gamma_1^{(1)}(\Gamma_1^{(1)}+\Gamma_2^{(2)})]\\
 &\hspace{0.57cm}= \frac{i\mu}{8}(Q_1Q_{1xx}^{*}+\mu^2Q_1^2Q_1^{*2}+\mu^2Q_1Q_2Q_1^{*}Q_2^{*}),\\
 \lambda^{-k}:&\Gamma_1^{(k+1)}=-\frac{i}{2}[-\frac{Q_{1x}}{Q_1}\Gamma_1{(k)}+\Gamma_{1x}^{(k)}+\mu\sum^{k-1}_{j=1}\Gamma_1^{(j)}(\Gamma_1^{(k-j)}+\Gamma_2^{(k-j)})]~(k=3,4,5,\dots),\label{xt-3-6d}
 \end{align}
 \end{subequations}
 and
 \begin{subequations}\label{xt-3-7}
 \begin{align}
 \lambda^{0}:&\Gamma_2^{(1)}=-\frac{i}{2}\mu Q_2Q_2^*,\\
 \lambda^{-1}:&\Gamma_2^{(2)}=-\frac{i}{2}(\Gamma_{2x}^{(1)}-\frac{Q_{2x}}{Q_2}\Gamma_2^{(1)})=-\frac{1}{4}\mu Q_2Q_{2x}^*,\\
 \lambda^{-2}:&\nonumber\Gamma_2^{(3)}=\frac{i\mu}{8}[-\frac{Q_{2x}}{Q_2}\Gamma_2^{(2)}+\Gamma_{2x}^{(2)}+\mu\Gamma_2^{(1)}(\Gamma_1^{(1)}+\Gamma_2^{(1)})]\\
 &\hspace{0.57cm}=\frac{i\mu}{8}(Q_2Q_{2xx}^{*}+\mu^2Q_2^2Q_2^{*2}+\mu^2Q_1Q_2Q_1^{*}Q_2^{*}),\\
 \lambda^{-k}:&\Gamma_2^{(k+1)}=-\frac{i}{2}[-\frac{Q_{2x}}{Q_2}\Gamma_2^{(k)}+\Gamma_{2x}^{(k)}+\mu\sum^{k-1}_{j=1}\Gamma_2^{(j)}(\Gamma_1^{(k-j)}+\Gamma_2^{(k-j)})]~(k=3,4,5,\dots).\label{xt-3-7d}
\end{align}
 \end{subequations}
 In order to use the associated evolution equation Eq.(\ref{xt-3-3}), we consider the compatibility condition $(\dfrac{\psi_{1x}}{\psi_1})_t=(\dfrac{\psi_{1t}}{\psi_1})_x$. Using Eqs.(\ref{xt-3-6}) and Eqs.(\ref{xt-3-7}), then equating the coefficients of the same power of $\lambda$ again, we can get the infinitely-many conservation laws for Eq.(\ref{xt-3-1})
 \begin{equation}
 \dfrac{\partial U_j}{\partial t}=\dfrac{\partial F_j}{\partial x}~(j=1,2,3,\dots),
 \end{equation}
 where $U_j$'s and $F_j$'s refer to conserved densities and conserved fluxes respectively.

 The first  conservation law
 \begin{subequations}
 \begin{align}
&U_1=\Gamma_1^{(1)}+\Gamma_2^{(1)}=-\frac{i\mu}{2}(|Q_1|^2+|Q_2|^2),\\
&\nonumber F_1=2(\Gamma_1^{(2)}+\Gamma_2^{(2)})+i(\frac{Q_{1x}}{Q_1}\Gamma_1^{(1)}+\frac{Q_{2x}}{Q_2}\Gamma_2^{(1)})-2\beta x(\Gamma_1^{(1)}+\Gamma_2^{(1)})\\
&\hspace{0.5cm}=\frac{\mu}{2}[Q_1^{*}Q_{1x}+Q_2^{*}Q_{2x}-Q_1Q_{1x}^{*}-Q_2Q_{2x}^{*}+2i\beta x(|Q_1|^2+|Q_2|^2)],
 \end{align}
 \end{subequations}
 the second conservation law
 \begin{subequations}
 \begin{align}
&U_2=\Gamma_1^{(2)}+\Gamma_2^{(2)}=-\frac{\mu}{4}(Q_1Q_{1x}^{*}+Q_2Q_{2x}^{*}),\\
&\nonumber F_2=2(\Gamma_1^{(3)}+\Gamma_2^{(3)})+i(\frac{Q_{1x}}{Q_1}\Gamma_1^{(2)}+\frac{Q_{2x}}{Q_2}\Gamma_2^{(2)})-2\beta x(\Gamma_1^{(2)}+\Gamma_2^{(2)})\\
&\hspace{0.5cm}=\frac{i\mu}{4}[Q_1Q_{1xx}^{*}+Q_2Q_{2xx}^{*}+\mu^2(|Q_1|^2+|Q_2|^2)^2-\frac{i}{2}\beta x(Q_1Q_{1x}^{*}+Q_2Q_{2x}^{*})-Q_{1x}Q_{1x}^{*}-Q_{2x}Q_{2x}^{*}],
 \end{align}
 \end{subequations}
 the k-th conservation law
 \begin{subequations}
 \begin{align}
 &U_k=\Gamma_1^{(k)}+\Gamma_2^{(k)}~(k\geq3),\\
 &\nonumber F_k=2(\Gamma_1^{(k+1)}+\Gamma_2^{(k+1)})+i(\frac{Q_{1x}}{Q_1}\Gamma_1^{(k)}+\frac{Q_{2x}}{Q_2}\Gamma_2^{(k)})-2\beta x(\Gamma_1^{(k)}+\Gamma_2^{(k)})\\
 &\hspace{0.5cm}=i[2(\frac{Q_{1x}}{Q_1}\Gamma_1^{(k)}+\frac{Q_{2x}}{Q_2}\Gamma_2^{(k)})-(\Gamma_{1x}^{(k)}+\Gamma_{2x}^{(k)})-\mu\sum_{j=1}^{k-1}(\Gamma_1^{(j)}\Gamma_1^{(k-j)}+2\Gamma_1^{(j)}\Gamma_2^{(k-j)}
 +\Gamma_2^{(j)}\Gamma_2^{(k-j)})+2i\beta x(\Gamma_1^{(k)}+\Gamma_2^{(k)})],
 \end{align}
 \end{subequations}
 where $\Gamma_1^{(k)}$ and $\Gamma_2^{(k)}$ can be directly derived from Eqs. (\ref{xt-3-6d}) and (\ref{xt-3-7d}).

\section{Darboux transformation and nonautonomous soliton solutions}  
In this section,  we will construct Darboux transformation of Lax pair (\ref{xt-3-2}) and (\ref{xt-3-3}) with nonisospectral parameter $\lambda$. Gu found a unified approach to construct Darboux transformations for  isospectral AKNS (Ablowitz-Kaup-Newell-Segur) hierarchy\cite{3-31}, and it had been employed in many integrable equations\cite{3-9,3-37,3-38,3-39}. However, there are also other nonisospectral integrable systems\cite{3-19,3-27}, a method to construct Darboux transformation for a class of nonisospectral cases \cite{3-40} was utilized by Ci\'{e}sl\'{i}nski, which is greatly different from Gu's method. After this, Zhou \cite{3-32,3-33} generalized Gu's formula to the nonisospectral case. Here, we use Zhou's method to construct Darboux transformation about Eq.(\ref{xt-3-1}).

Let $\Psi=(\psi[1](\xi_j),\phi[1](\xi_j),\chi[1](\xi_j))^T$ be a special vector solution of Lax pair (\ref{xt-3-2}) and (\ref{xt-3-3}), with choosing the seed solution of Eq.(\ref{xt-3-1}) ~$q_1 = q_1[0],~ q_2 = q_2[0]$ at ~$\lambda =\lambda_j=\dfrac{\alpha}{4\beta}+\xi_j e^{-2\beta t}$.

Then, we give the first-step Darboux transformation of Eq.(\ref{xt-3-1})
\begin{eqnarray}
&&\Psi[1]=T[1]\Psi,\quad T[1]=\rho_1(\lambda)(\lambda I-H[1]\Lambda_1H[1]^{-1}),\label{xt-3-12}\\
&&q_1[1]=q_1[0]+\dfrac{2i(\lambda_1^{*}-\lambda_1)\psi[1](\xi_1) \phi[1](\xi_1)^{*}}{\mu(|\psi[1](\xi_1)|^2+|\phi[1](\xi_1)|^2+|\chi[1](\xi_1)|^2)}e^{\dfrac{i\beta x^2}{2}},\label{xt-3-8}\\
&&q_2[1]=q_2[0]+\dfrac{2i(\lambda_1^{*}-\lambda_1)\psi[1](\xi_1) \chi[1](\xi_1)^{*}}{\mu(|\psi[1](\xi_1)|^2+|\phi[1](\xi_1)|^2+|\chi[1](\xi_1)|^2)}e^{\dfrac{i\beta x^2}{2}},\label{xt-3-9}
\end{eqnarray}
where $I$ is the $3\times3$ identity matrix,

\begin{eqnarray}
&&\nonumber\rho_1(\lambda)=[det(\lambda I-H[1]\Lambda_1H[1]^{-1})]^{-\frac{1}{3}}\\
&&\hspace{0.84cm}=[(\lambda-\lambda_1)(\lambda-\lambda_1^{*})^2]^{-\frac{1}{3}},\\
&&H[1]=\begin{bmatrix}\psi[1](\xi_1)&\phi[1](\xi_1)^{*}&0\\
                       \phi[1](\xi_1)&-\psi[1](\xi_1)^{*}&\chi[1](\xi_1)^{*}\\
                       \chi[1](\xi_1)&0&-\phi[1](\xi_1)^{*} \end{bmatrix},\quad
\Lambda_1=\begin{bmatrix} \lambda_1&0&0\\ 0&\lambda_1^{*}&0\\ 0&0&\lambda_1^{*} \end{bmatrix}.
\end{eqnarray}
The above DT should hold the two relations $T[1]_x+T[1]U=U[1]T[1]$ ~and~ $T[1]_t+T[1]V=V[1]T[1]$, where $U[1]$ and $V[1]$ enjoys the same forms as $U$ and $V$ except that $(q_1[0], q_2[0])$ is replaced with $(q_1[1],q_2[1])$. The accuracy of these two relations had been verified by us though Maple software.

Similarly, the first-step Darboux transformation for one-component G-P equation can be expressed as
\begin{eqnarray}
&&\widetilde{\Psi}=T\Psi_0,\quad T=\rho(\lambda)(\lambda I-H\Lambda_1H^{-1}),\\
&&q_1=q_0+\dfrac{2i(\lambda_1^{*}-\lambda_1)\psi(\xi_1) \phi(\xi_1)^{*}}{\mu(|\psi(\xi_1)|^2+|\phi(\xi_1)|^2}e^{\dfrac{i\beta x^2}{2}}\label{xt-3-33},
\end{eqnarray}
where $I$ is the $2\times2$ identity matrix and $\Psi_0=(\psi(\xi_1),\phi(\xi_1))^T$,
\begin{eqnarray}
&&\nonumber\rho(\lambda)=[(\lambda-\lambda_1)(\lambda-\lambda_1^{*})]^{-\frac{1}{2}},\\
&&H=\begin{bmatrix}\psi(\xi_1)&\phi(\xi_1)^{*}\\
                       \phi(\xi_1)&-\psi(\xi_1)^{*}\end{bmatrix},\quad
\Lambda_1=\begin{bmatrix} \lambda_1&0\\ 0&\lambda_1^{*} \end{bmatrix}.
\end{eqnarray}
 In \cite{3-19} and \cite{3-24}, $\rho(\lambda)$ was not included in the Darboux transformation of one-component G-P equation, which brought about their DT didn't admit the t-part of Lax pair, without iterating DT, the expression of the first-step solution $q_1$ is right. In \cite{3-19}, based on their incorrect DT, the two solitons are also not correct, besides, in \cite{3-19} and \cite{3-24}, the second-order rogue waves are all not reasonable via the generalized DT \cite{3-39}. It is necessary for us to reconstruct the DT for Eq.(\ref{xt-3-1}) and its multi-component generalization.

The second-step Darboux transformation
\begin{eqnarray}
&&\Psi[2]=T[2]T[1]\Psi,\quad T[2]=\rho_2(\lambda)(\lambda I-H[2]\Lambda_2H[2]^{-1}),\\
&&q_1[2]=q_1[1]+\dfrac{2i(\lambda_2^{*}-\lambda_2)\psi[2](\xi_2) \phi[2](\xi_2)^{*}}{\mu(|\psi[2](\xi_2)|^2+|\phi[2](\xi_2)|^2+|\chi[2](\xi_2)|^2)}e^{\dfrac{i\beta x^2}{2}},\label{xt-3-13}\\
&&q_2[2]=q_2[1]+\dfrac{2i(\lambda_2^{*}-\lambda_2)\psi[2](\xi_2) \chi[2](\xi_2)^{*}}{\mu(|\psi[2](\xi_2)|^2+|\phi[2](\xi_2)|^2+|\chi[2](\xi_2)|^2)}e^{\dfrac{i\beta x^2}{2}},\label{xt-3-14}
\end{eqnarray}
with
\begin{eqnarray}
&& T[1]|_{\lambda=\lambda_2}(\psi[1](\xi_2),\phi[1](\xi_2),\chi[1](\xi_2))^T=(\psi[2](\xi_2),\phi[2](\xi_2),\chi[2](\xi_2))^T,\\
&&\rho_2(\lambda)=[(\lambda-\lambda_2)(\lambda-\lambda_2^{*})^2]^{-\frac{1}{3}},\\
&&H[1]=\begin{bmatrix}\psi[1](\xi_2)&\phi[1](\xi_2)^{*}&0\\
                       \phi[1](\xi_2)&-\psi[1](\xi_2)^{*}&\chi[1](\xi_2)^{*}\\
                       \chi[1](\xi_2)&0&-\phi[1](\xi_2)^{*} \end{bmatrix},\quad
\Lambda_2=\begin{bmatrix} \lambda_2&0&0\\ 0&\lambda_2^{*}&0\\ 0&0&\lambda_2^{*} \end{bmatrix}.
\end{eqnarray}

The \emph{N}-step Darboux transformation
\begin{eqnarray}
&&\Psi[N]=T[N]T[N-1]\cdots T[2]T[1]\Psi,\quad T[N]=\rho_N(\lambda)(\lambda I-H[N]\Lambda_NH[N]^{-1}),\\
&&q_1[N]=q_1[N-1]+\dfrac{2i(\lambda_N^{*}-\lambda_N)\psi[N-1](\xi_{N-1}) \phi[N-1](\xi_{N-1})^{*}}{\mu(|\psi[N-1](\xi_{N-1})|^2+|\phi[N-1](\xi_{N-1})|^2+|\chi[N-1](\xi_{N-1})|^2)}e^{\dfrac{i\beta x^2}{2}},\\
&&q_2[N]=q_2[N-1]+\dfrac{2i(\lambda_N^{*}-\lambda_N)\psi[N-1](\xi_{N-1}) \chi[N-1](\xi_{N-1})^{*}}{\mu(|\psi[N-1](\xi_{N-1})|^2+|\phi[N-1](\xi_{N-1})|^2+|\chi[N-1](\xi_{N-1})|^2)}e^{\dfrac{i\beta x^2}{2}},
\end{eqnarray}
with
\begin{eqnarray}
&&\nonumber T[N-2]|_{\lambda=\lambda_{N-1}}T[N-3]|_{\lambda=\lambda_{N-2}}\cdots T[2]|_{\lambda=\lambda_3}T[1]|_{\lambda=\lambda_2}(\psi[1](\xi_{N-1}),\phi[1](\xi_{N-1}),\chi[1](\xi_{N-1}))^T\\
&&\hspace{0.84cm}=(\psi[N-1](\xi_{N-1}),\phi[N-1](\xi_{N-1}),\chi[N-1](\xi_{N-1}))^T,\label{xt-3-29}\\
&&\rho_j(\lambda)=[(\lambda-\lambda_j)(\lambda-\lambda_j^{*})^2]^{-\frac{1}{3}},\\
&&H[j]=\begin{bmatrix}\psi[1](\xi_j)&\phi[1](\xi_j)^{*}&0\\
                       \phi[1](\xi_j)&-\psi[1](\xi_j)^{*}&\chi[1](\xi_j)^{*}\\
                       \chi[1](\xi_j)&0&-\phi[1](\xi_j)^{*} \end{bmatrix},\quad
\Lambda_j=\begin{bmatrix} \lambda_j&0&0\\ 0&\lambda_j^{*}&0\\ 0&0&\lambda_j^{*} \end{bmatrix}~(3\leq j\leq N),
\end{eqnarray}
where $\lambda_i\neq\lambda_j~(i\neq j)$. Due to the $\rho_j=[(\lambda-\lambda_j)(\lambda-\lambda_j^{*})^2]^{-\frac{1}{3}}~(j=1,2,3,\dots,N)$ exist in the j-step DT of Eq.(\ref{xt-3-1}), here, we can not give the determinant representations of DT and solutions.

In the following contents, we will give nonautonomous solitons of Eq.(\ref{xt-3-1}) though the above DT. For convenience, we chose the trivial seed solution $q_1[0]=0$ and $q_2[0]=0$, with $\xi=\xi_j~(j=1,2,3,\dots,N)$. Thus, the special vector solutions of Lax pair Eqs.(\ref{xt-3-2}) and (\ref{xt-3-3}) can be given
\begin{equation}\label{xt-3-10}
\psi[1](\xi_j)=e^{\theta_j},\quad\phi[1](\xi_j)=e^{-\theta_j},\quad \chi[1](\xi_j)=-2e^{-\theta_j},
\end{equation}
where
\begin{equation*}
\theta_j=-i(\frac{\alpha}{4\beta}+\xi_je^{-2\beta t})x+\frac{i}{8\beta^2}(-\alpha^2t+4\xi_j^2\beta e^{-4\beta t}+4\alpha\xi_je^{-2\beta t}).
\end{equation*}
In order to get one nonautonomous soliton for Eq.(\ref{xt-3-1}), we substitute Eq.(\ref{xt-3-10}) into Eqs.(\ref{xt-3-8}) and (\ref{xt-3-9}), and the expressions of one soliton can be given

\begin{eqnarray}
&& q_1[1]=\dfrac{2i(\lambda_1^{*}-\lambda_1)\psi[1](\xi_1) \phi[1](\xi_1)^{*}}{\mu(|\psi[1](\xi_1)|^2+|\phi[1](\xi_1)|^2+|\chi[1](\xi_1)|^2)}e^{\dfrac{i\beta x^2}{2}}=\dfrac{2i(\xi_1^{*}-\xi_1)e^{\theta_1-\theta_1^{*}}}{\mu(e^{\theta_1+\theta_1^{*}}+5e^{-\theta_1-\theta_1^{*}})}e^{\dfrac{i\beta x^2}{2}-2\beta t},\\
&&q_2[1]=\dfrac{2i(\lambda_1^{*}-\lambda_1)\psi[1](\xi_1) \chi[1](\xi_1)^{*}}{\mu(|\psi[1](\xi_1)|^2+|\phi[1](\xi_1)|^2+|\chi[1](\xi_1)|^2)}e^{\dfrac{i\beta x^2}{2}}
=-\dfrac{4i(\xi_1^{*}-\xi_1)e^{\theta_1-\theta_1^{*}}}{\mu(e^{\theta_1+\theta_1^{*}}+5e^{-\theta_1-\theta_1^{*}})}e^{\dfrac{i\beta x^2}{2}-2\beta t}.
\end{eqnarray}
Then the modules of $q_1[1]$ and $q_2[1]$ can also be expressed as
\begin{eqnarray}
&&|q_1[1]|=\dfrac{4e^{-2\beta t}|Im(\xi_1)|}{|\mu|\left(e^{-\dfrac{e^{-2\beta t}Im(\xi_1)(2e^{-2\beta t} Re(\xi_1)\beta-2x\beta^2+\alpha)}{\beta^2}}+5e^{\dfrac{e^{-2\beta t}Im(\xi_1)(2e^{-2\beta t} Re(\xi_1)\beta-2x\beta^2+\alpha)}{\beta^2}}\right)},\\
&&|q_2[1]|=\dfrac{8e^{-2\beta t}|Im(\xi_1)|}{|\mu|\left(e^{-\dfrac{e^{-2\beta t}Im(\xi_1)(2e^{-2\beta t} Re(\xi_1)\beta-2x\beta^2+\alpha)}{\beta^2}}+5e^{\dfrac{e^{-2\beta t}Im(\xi_1)(2e^{-2\beta t} Re(\xi_1)\beta-2x\beta^2+\alpha)}{\beta^2}}\right)},
\end{eqnarray}
where $\xi_1=Re(\xi_1)+iIm(\xi_1)$, then we set the following equality
\begin{equation}\label{xt-3-11}
2e^{-2\beta t} Re(\xi_1)\beta-2x\beta^2+\alpha=0.
\end{equation}

From the above Eq.(\ref{xt-3-11}), the propagation velocity of one soliton of Eq.(\ref{xt-3-1}) can be derived
\begin{equation*}
v=\frac{\mathrm{d}x}{\mathrm{d}x}=-2Re(\xi_1)e^{-2\beta t},
\end{equation*}
it can be found that $v$ is only depend on $t$ and $Re(\xi_1)$. When $\beta>0$, the absolute value of $v$ decays exponentially as $t$ increases, while increases exponentially as $t$ increases if $\beta<0$ . And $Re(\xi_1)$ determines both  propagation direction of one soliton and the value of $v$. When $Re(\xi_1)$=0, the propagation velocity of one soliton is zero, which indicates this  soliton is stationary, as can be seen from Fig.1. When $Re(\xi_1)>0$, the one nonautonomous soliton propagates along the negative direction of x-axis; if $Re(\xi_1)<0$, it propagates along the positive direction of x-axis. These interesting results can be found in Fig.2 and Fig.3 respectively. With the value of $|\mu|$ increasing, the amplitude of one soliton decreases. From Fig.3 and Fig.4, we can observe that the parameter $\alpha$ affects the localization of one soliton.

Through Eqs.(\ref{xt-3-12}), (\ref{xt-3-13}) and (\ref{xt-3-14}), the two nonautonomous solitons of Eq.(\ref{xt-3-1}) can be expressed
\begin{eqnarray}
&&q_1[2]=\frac{2i}{\mu}\left(\dfrac{(\xi_1^{*}-\xi_1)e^{\theta_1-\theta_1^{*}}}{e^{\theta_1+\theta_1^{*}}+5e^{-\theta_1-\theta_1^{*}}}+\dfrac{(\xi_2^{*}-\xi_2)G_1}{F}\right)e^{\dfrac{i\beta x^2}{2}-2\beta t},\label{xt-3-15}\\
&&q_2[2]=\frac{2i}{\mu}\left(-2\dfrac{(\xi_1^{*}-\xi_1)e^{\theta_1-\theta_1^{*}}}{e^{\theta_1+\theta_1^{*}}+5e^{-\theta_1-\theta_1^{*}}}+\dfrac{(\xi_2^{*}-\xi_2)G_2}{F}\right)e^{\dfrac{i\beta x^2}{2}-2\beta t},\label{xt-3-16}
\end{eqnarray}
where
\begin{eqnarray*}
&&G_1=5(-{\xi_1^{*}}^2+\xi_1^{*}\xi_1+\xi_1^{*}\xi_2-\xi_1\xi_2)e^{-2\theta_1^{*}+\theta_2^{*}+\theta_2}+5(\xi_1^{*}\xi_2^{*}-2\xi_1^{*}\xi_1-\xi_1^{*}\xi_2+\xi_2^{*}\xi_1+\xi_1\xi_2)e^{-\theta_2^{*}+\theta_2}\\
&&\hspace{0.5cm}+25({\xi_1^{*}}^2-\xi_1^{*}\xi_2^{*}-\xi_1^{*}\xi_1+\xi_2^{*}\xi_1)e^{-\theta_2^{*}-2\theta_1^{*}-\theta_2}+(-\xi_1^{*}\xi_1-\xi_1^{*}\xi_2+\xi_1^2+\xi_1\xi_2)e^{2\theta_1+\theta_2^{*}+\theta_2}\\
&&\hspace{0.5cm}+5({\xi_1^{*}}^2-2\xi_1^{*}\xi_1+\xi_1^2)e^{2\theta_1-2\theta_1^{*}+\theta_2^{*}-\theta_2}+25(-{\xi_1^{*}}^2+\xi_1^{*}\xi_2^{*}+\xi_1^{*}\xi_2-\xi_2^{*}\xi_2)e^{-2\theta_1-\theta_2^{*}-2\theta_1^{*}+\theta_2}\\
&&\hspace{0.5cm}+(\xi_2^{*}\xi_1+\xi_2^{*}\xi_2-\xi_1^2-\xi_1\xi_2)e^{2\theta_1+2\theta_1^{*}-\theta_2^{*}+\theta_2}+5(-\xi_1^{*}\xi_2^{*}+\xi_1^{*}\xi_1+\xi_2^{*}\xi_1-\xi_1^2)e^{2\theta_1-\theta_2^{*}-\theta_2},\\
&&G_2=10({\xi_1^{*}}^2-\xi_1^{*}\xi_1-\xi_1^{*}\xi_2+\xi_1\xi_2)e^{-2\theta_1^{*}+\theta_2^{*}+\theta_2}+2(\xi_1^{*}\xi_1+\xi_1^{*}\xi_2-\xi_1^2-\xi_1\xi_2)e^{2\theta_1+\theta_2^{*}+\theta_2}\\
&&\hspace{0.5cm}+10(-{\xi_1^{*}}^2+2\xi_1^{*}\xi_1-\xi_1^2)e^{2\theta_1-2\theta_1^{*}+\theta_2^{*}-\theta_2}+50(-{\xi_1^{*}}^2+\xi_1^{*}\xi_2^{*}+\xi_1^{*}\xi_1-\xi_2^{*}\xi_1)e^{-\theta_2^{*}-2\theta_1^{*}-\theta_2}\\
&&\hspace{0.5cm}+50({\xi_1^{*}}^2-\xi_1^{*}\xi_2^{*}-\xi_1^{*}\xi_2+\xi_2^{*}\xi_2)e^{-2\theta_1-\theta_2^{*}-2\theta_1^{*}+\theta_2}+10(-\xi_1^{*}\xi_1+\xi_1^{*}\xi_2+\xi_1^2-\xi_1\xi_2)e^{2\theta_1-\theta_2^{*}-\theta_2}\\
&&\hspace{0.5cm}+10(2\xi_1^{*}\xi_1-\xi_2^{*}\xi_1-\xi_2^{*}\xi_2-\xi_1\xi_2+\xi_2^2)e^{-\theta_2^{*}+\theta_2}+2(\xi_1^2-\xi_2^2)e^{2\theta_1+2\theta_1^{*}-\theta_2^{*}+\theta_2},\\
&&F=25(-\xi_1^{*}\xi_2^{*}+\xi_1^{*}\xi_2+\xi_2^{*}\xi_1-\xi_1\xi_2)e^{-2\theta_1^{*}-\theta_2+\theta_2^{*}}-(\xi_1^{*}\xi_1+\xi_1^{*}\xi_2+\xi_2^{*}\xi_1+\xi_2^{*}\xi_2)e^{2\theta_1+2\theta_1^{*}+\theta_2+\theta_2^{*}}\\
&&\hspace{0.5cm}+125(-\xi_1^{*}\xi_1+\xi_1^{*}\xi_2+\xi_2^{*}\xi_1-\xi_2^{*}\xi_2)e^{-2\theta_1-\theta_2^{*}-2\theta_1^{*}-\theta_2}+(\xi_1^{*}\xi_2^{*}-5\xi_1^{*}\xi_1+4\xi_1^{*}\xi_2-\xi_2^{*}\xi_2+5\xi_1\xi_2-4\xi_2^2)e^{2\theta_1+2\theta_1^{*}-\theta_2^{*}-\theta_2}\\
&&\hspace{0.5cm}+5(5\xi_1\xi_2-10\xi_1^{*}\xi_1+5\xi_1^{*}\xi_2+\xi_1^{*}\xi_1-6\xi_2^{*}\xi_2+9\xi_1\xi_2-4\xi_2^2)e^{-\theta_2-\theta_2^{*}}+25(-\xi_1^{*}\xi_2^{*}+\xi_1^{*}\xi_2+\xi_2^{*}\xi_1-\xi_1\xi_2)e^{-2\theta_1+\theta_2-\theta_2^{*}}\\
&&\hspace{0.5cm}+(-\xi_1^{*}\xi_2^{*}-9\xi_1^{*}\xi_2+\xi_2^{*}\xi_1+9\xi_1\xi_2)e^{2\theta_1^{*}+\theta_2-\theta_2^{*}}+5(\xi_1^{*}\xi_2^{*}+\xi_1^{*}\xi_2-\xi_2^{*}\xi_1-\xi_1\xi_2)e^{2\theta_1-\theta_2+\theta_2^{*}}+25(\xi_1^{*}\xi_2^{*}-\xi_1^{*}\xi_1-\xi_2^{*}\xi_2+\\
&&\hspace{0.5cm}\xi_1\xi_2)e^{-2\theta_1-2\theta_1^{*}+\theta_2+\theta_2^{*}}+5(-\xi_1^{*}\xi_2^{*}-2\xi_1^{*}\xi_1+\xi_1^{*}\xi_2+\xi_2^{*}\xi_1+2\xi_2^{*}\xi_2-\xi_1\xi_2)e^{\theta_2+\theta_2^{*}}.
\end{eqnarray*}

From the above two equations (\ref{xt-3-15}) and (\ref{xt-3-16}), we can find that the parameter $\mu$ affects amplitude of the two nonautonomous solitons. Similarly, the velocity of soliton can be written as $v_j=-2Re({\xi_j})e^{-2\beta t}~(j=1,2)$. When $v_1\neq v_2~(Re(\xi_1)\neq Re(\xi_2))$, the two nonautonomous solitons will interact with each other, which include overtaking and head-on interaction. This phenomen can be seen in Fig.5. Setting $v_1=v_2~(Re(\xi_1)=Re(\xi_2))$, the bound state\cite{3-41} in two  nonautonomous solitions  occur in Fig.6. Meanwhile, by iterating $\emph{N}$ times of DT, we can get $\emph{N}$-nonautonomous solitons of Eq.(\ref{xt-3-1}).
%
\section{Nonautonomous breather and rogue wave}
In this section, we will present the breather and rogue wave solutions for Eq.(\ref{xt-3-1}) through the above DT. In order to obtain these rational solutions, we should consider the following seed solutions with nonzero background and the amplitudes of background waves decay exponentially as $t$ increases, so this is greatly different from those rational solutions in autonomous systems, 
\begin{equation}
q_1[0]=d_1e^{-2\beta t+i(\tfrac{\beta}{2}x^2+\theta)},\quad q_2[0]=d_2e^{-2\beta t+i(\tfrac{\beta}{2}x^2+\theta)},\label{xt-3-17}
\end{equation}
where
\begin{equation}
\theta=(-\frac{\alpha}{2\beta}+ce^{-2\beta t})x-\frac{1}{4\beta^2}e^{-4\beta t}[2\beta\mu^2(d_1^2+d_2^2)-\beta c^2+2\alpha ce^{2\beta t}+\alpha^2e^{4\beta t}t],\label{xt-3-21}
\end{equation}
here, $d_1,~d_2,~c$ are all real constants, and $d_1\neq d_2$. Then, choosing the spectral parameter $\lambda=\lambda_1=\dfrac{\alpha}{4\beta}+\xi_1e^{-2\beta t}$ and substituting (\ref{xt-3-17}) into Lax pair (\ref{xt-3-2}) and (\ref{xt-3-3}), we can get the special vector solution of Lax  pair
\begin{eqnarray}
\Psi_1=\begin{pmatrix}\dfrac{i}{2}\left[c_2(\beta c+2\xi_1\beta-\eta)e^{m_2}+c_3(\beta c+2\xi\beta+\eta)e^{m_3}\right]e^{\tfrac{2i}{3}\theta}\\
                       \left[-c_1d_2e^{m_1}+\mu d_1\beta(c_2e^{m_2}+c_3e^{m_3})\right]e^{-\tfrac{i}{3}\theta}\\
                       \left[c_1d_1e^{m_1}+\mu d_2\beta(c_2e^{m_2}+c_3e^{m_3})\right]e^{-\tfrac{i}{3}\theta}\end{pmatrix},\label{xt-3-18}
\end{eqnarray}
where
\begin{eqnarray*}
\nonumber &&\eta=\sqrt{\beta^2[4\mu^2(d_1^2+d_2^2)+(c+2c\xi_1)^2]},\\
\nonumber &&m_1=[(\frac{i}{3}c+i\xi_1)e^{-2\beta t}+\frac{i\alpha}{12\beta}]x-\frac{i\alpha}{6\beta^2}(3\xi_1+c)e^{-2\beta t}+\frac{i}{12\beta}[-2\mu^2(d_1^2+d_2^2)+c^2-6\xi_1^2]e^{-4\beta t}+\frac{i\alpha^2}{24\beta^2}t,\\
\nonumber &&m_2=[-\frac{i}{6}(c-\frac{3\eta}{\beta})e^{-2\beta t}+\frac{i\alpha}{12\beta}]x+\frac{i\alpha}{12\beta^2}(\beta c-3\eta)e^{-2\beta t}+\frac{i}{24\beta^2}[2\beta\mu^2(d_1^2+d_2^2)-\beta c^2+3\eta c-6\eta \xi_1]e^{-4\beta t}\\
\nonumber&&\hspace{0.6cm}+\frac{i\alpha^2}{24\beta^2}t,\\
\nonumber &&m_3=[-\frac{i}{6}(c+\frac{3\eta}{\beta})e^{-2\beta t}+\frac{i\alpha}{12\beta}]x+\frac{i\alpha}{12\beta^2}(\beta c+3\eta)e^{-2\beta t}+\frac{i}{24\beta^2}[2\beta\mu^2(d_1^2+d_2^2)-\beta c^2-3\eta c+6\eta \xi_1]e^{-4\beta t}\\
\nonumber&&\hspace{0.6cm}+\frac{i\alpha^2}{24\beta^2}t,\\
\end{eqnarray*}
with $c_1,~c_2,~c_3$ being all arbitrary real constants. In order to construct nonautonomous breather, we should require $\eta$ is a pure imaginary number. Choosing $\xi_1=-\dfrac{c}{2}+i\mu Im(\xi_1)$ with $Im(\xi_1)$ being a real constant, the expression of $\eta$ can be simplified as $\eta=2\sqrt{\beta^2\mu^2(d_1^2+d_2^2-Im(\xi_1)^2)}$, and $Im(\xi_1)>d_1^2+d_2^2$. Substituting Eqs.(\ref{xt-3-17}) and (\ref{xt-3-18}) into Eqs.(\ref{xt-3-8}) and (\ref{xt-3-9}) respectively, the compact expressions of one breather can be constructed
\begin{eqnarray}
&&q_1[1]=(d_1+\dfrac{4Im(\xi_1)H_1}{H_3})e^{-2\beta t+i(\tfrac{\beta}{2}x^2+\theta)},\label{xt-3-19}\\
&&q_2[1]=(d_2+\dfrac{4Im(\xi_1)H_2}{H_3})e^{-2\beta t+i(\tfrac{\beta}{2}x^2+\theta)},\label{xt-3-20}
\end{eqnarray}
where
\begin{eqnarray*}
&&H_1=-(\mu\beta Im({\xi_1})+\eta_0)(\mu d_1\beta c_2c_3e^{m_2+m_3}+\mu \beta d_1c_3^2e^{m_3+m_3^{*}}-d_2c_1c_3e^{m_1^{*}+m_3})+(\eta_0-\mu\beta Im({\xi_1}))(\mu d_1\beta c_2c_3e^{m_2+m_3^{*}}\\
&&\hspace{0.6cm}+\mu d_1\beta c_2^2e^{2m_2}-d_2c_1c_2e^{m_1^{*}+m_2}),\\
&&H_2=-(\mu\beta Im(\xi_1)+\eta_0)(\mu d_2\beta c_2c_3e^{m_2^{*}+m_3}+\mu d_2\beta c_3^2e^{m_3+m_3^{*}}+d_1 c_1c_3e^{m_1^{*}+m_3})+(\eta_0-\mu\beta Im(\xi_1))(\mu d_2\beta c_2c_3e^{m_3+m_3^{*}}\\
&&\hspace{0.6cm}+\mu d_2\beta c_2^2e^{m_2+m_2^{*}}+d_1c_1c_2e^{m_1^{*}+m_2}),\\
&&H_3=c_1^2(d_1^2+d_2^2)e^{m_1+m_1^{*}}+\mu c_2^2\beta(2\mu\beta Im(\xi_1)^2-\mu d_1^2\beta-2Im(\xi_1)\eta_0)e^{m_2+m_2^{*}}+2\mu \beta c_3^2 Im(\xi_1)(\mu\beta Im(\xi_1)+\eta_0)e^{m_3+m_3^{*}}\\
&&\hspace{0.6cm}+2\mu^2\beta^2c_2c_3(d_1^2+d_2^2)e^{m_2+m_3^{*}}+\mu^2\beta^2c_2c_3(d_1^2+2d_2^2)e^{m_2^{*}+m_3}+\mu^2d_1^2\beta^2c_2^2e^{2m_2}-\mu\beta d_1d_2c_1c_2e^{m_1+m_2}\\
&&\hspace{0.6cm}+\mu^2d_1^2\beta^2c_2c_3e^{m_2+m_3}+\mu\beta d_1d_2c_1c_2e^{m_1+m_2^{*}},
\end{eqnarray*}
with $\eta_0=\sqrt{\beta^2\mu^2(Im(\xi_1)^2-d_1^2-d_2^2)}$. From Eqs.(\ref{xt-3-19}) and (\ref{xt-3-20}), we find that the breather of this coupled G-P equation has some deformations along the direction of $t$ due to the exponential decaying term, and the amplitude of the breather becomes small till being zero as $t$ increases. which can be seen in Fig.7. In other words, the breather in this couple G-P system has curved background, which is greatly determined by the parameter $\beta$. This type breather is very different from the autonomous system, such as nonlinear Schr\"{o}dinger equation\cite{3-41}, and this special breather also occurs in \cite{3-20}.


In order to obtain rogue wave of Eq.(\ref{xt-3-1}), we should choose the appropriate spectral $\xi_1$ and take the limit $\eta\rightarrow 0$.
For convenience, we set $\beta>0,~\mu>0,~c=0,~c_1=0,~\xi_1=ih$~($h$ is a real constant) in Eq.(\ref{xt-3-18}), here,$\eta=2i\beta\sqrt{h^2-\mu^2(d_1^2+d_2^2)}~(h>\sqrt{\mu^2(d_1^2+d_2^2)})$. Then choosing the appropriate values of $c_2$ and $c_3$, the special vector solution of Lax pair can be obtained
\begin{eqnarray}
\Psi_2=\begin{pmatrix}\left(r_2e^{A}-r_1e^{-A}\right)e^{\tfrac{2i}{3}\theta_0+B}\\
                       \rho_1\left(r_2e^{-A}-r_1e^{A}\right)e^{-\tfrac{i}{3}\theta_0+B}\\
                       \rho_2\left(r_2e^{-A}-r_1e^{A}\right)e^{-\tfrac{i}{3}\theta_0+B}\end{pmatrix},\label{xt-3-22}
\end{eqnarray}
where
\begin{eqnarray*}
&&r_1=\dfrac{\sqrt{h-\sqrt{h^2-\mu^2(d_1^2+d_2^2)}}}{\sqrt{h^2-\mu^2(d_1^2+d_2^2)}},\quad r_2=\dfrac{\sqrt{h+\sqrt{h^2-\mu^2(d_1^2+d_2^2)}}}{\sqrt{h^2-\mu^2(d_1^2+d_2^2)}},\\
&&\rho_1=\dfrac{d_1}{\sqrt{d_1^2+d_2^2}},\quad \rho_2=\dfrac{d_2}{\sqrt{d_1^2+d_2^2}},\quad A=\sqrt{h^2-\mu^2(d_1^2+d_2^2)}e^{-2\beta t}(x-\dfrac{\alpha+ih\beta e^{-2\beta t}}{2\beta^2}),\\
&&B=\frac{i}{24\beta^2}\left[2\alpha x+\alpha^2t+2\mu^2(d_1^2+d_2^2)e^{-4\beta t}\right],\quad \theta_0=-\frac{\alpha}{2\beta}x-\dfrac{e^{-4\beta t}}{4\beta^2}[2\beta\mu^2(d_1^2+d_2^2)+\alpha^2e^{4\beta t}t].
\end{eqnarray*}
In Eq.(\ref{xt-3-22}), taking the limit $h\rightarrow\sqrt{\mu^2(d_1^2+d_2^2)}$, we can get a new vector solution of Lax pair (\ref{xt-3-2}) and (\ref{xt-3-3})
\begin{eqnarray}
\Psi_3=\begin{pmatrix}(\dfrac{1}{\sqrt{\mu\sqrt{d_1^2+d_2^2}}}+\eta_1)e^{\theta_1}\\
                      \dfrac{1}{\sqrt{d_1^2+d_2^2}}(d_1\eta_1+\dfrac{1}{\sqrt{\mu\sqrt{d_1^2+d_2^2}}})e^{\theta_2}\\
                      \dfrac{1}{\sqrt{d_1^2+d_2^2}}(d_2\eta_1+\dfrac{1}{\sqrt{\mu\sqrt{d_1^2+d_2^2}}})e^{\theta_2}\end{pmatrix},\label{xt-3-23}
\end{eqnarray}
where
\begin{eqnarray*}
&&\eta_1=2\sqrt{\mu\sqrt{d_1^2+d_2^2}}e^{-2\beta t}(x-\dfrac{\alpha+i\sqrt{d_1^2+d_2^2}\mu\beta e^{-2\beta t}}{2\beta^2}),\\
&&\theta_1=-\frac{i}{24\beta^2}[2\mu^2(4\beta-1)(d_1^2+d_2^2)e^{-4\beta t}+3\alpha^2 t+8\alpha\beta x-2\alpha x],\\
&&\theta_2=\frac{i}{24\beta^2}[2\mu^2(2\beta+1)(d_1^2+d_2^2)e^{-4\beta t}+3\alpha^2 t+4\alpha \beta x+2\alpha x].
\end{eqnarray*}
Finally, substituting the eigenfunction (\ref{xt-3-23}) into Eqs.(\ref{xt-3-8}) and (\ref{xt-3-9}), the first-order nonautonomous rogue wave solution can be
derived in the following form. And we give some figures of this first-order rogue.
\begin{eqnarray}
&&q_1[1]=\frac{L_1}{L}e^{i\theta_3},\label{xt-3-24}\\
&&q_2[1]=\frac{L_2}{L}e^{i\theta_3},\label{xt-3-25}
\end{eqnarray}
with
\begin{eqnarray*}
&&\theta_3=\frac{\beta^2}{2}x^2-\frac{\alpha}{2\beta}x-\dfrac{e^{-4\beta t}}{4\beta^2}[2\beta\mu^2(d_1^2+d_2^2)+\alpha^2e^{4\beta t}t],\\
&&L=(4\beta^4\mu^2x^2d_1^2+4\beta^4\mu^2x^2d_2^2-4\alpha\beta^2\mu^2xd_1^2-4\alpha\beta^2\mu^2xd_2^2+\alpha^2\mu^2d_1^2+\alpha^2\mu^2d_2^2)e^{4\beta t}
+\mu^4d_1^4\beta^2+2\mu^4d_1^2d_2^2\beta^2\\
&&\hspace{0.6cm}+\mu^4d_2^4\beta^2+e^{8\beta t}\beta^4,\\
&&L_1=(4\beta^4\mu^2x^2d_1^3+4\beta^4\mu^2x^2d_1d_2^2-4\alpha\beta^2\mu^2xd_1^3-4\alpha\beta^2\mu^2xd_1d_2^2+\alpha^2\mu^2d_1^3+\alpha^2\mu^2d_1d_2^2)e^{4\beta t}+(-8d_1^3\beta^4\mu^3x^2\\
&&\hspace{0.6cm}-8d_1\beta^4\mu^3x^2d_2^2-4id_1^3\beta^3\mu^3-4id_1\beta^3\mu^3d_2^2+8d_1^3\alpha\beta^2\mu^3x+8d_1\alpha\beta^2\mu^3xd_2^2-2d_1^3\alpha^2\mu^3-2d_1\alpha^2\mu^3d_2^2)e^{2\beta t}\\
&&\hspace{0.6cm}-(2\beta^2\mu^5d_1^5+4\beta^2\mu^5d_1^3d_2^2+2\beta^2\mu^5d_1d_2^4)e^{-2\beta t}+2d_1\mu\beta^4e^{6\beta t}+d_1\beta^4e^{8\beta t}+d_1\mu^4d_2^4\beta^2+d_1^5\mu^4\beta^2+2d_1^3\mu^4d_2^2\beta^2,\\
&&L_2=(4\beta^4\mu^2x^2d_1^2d_2+4\beta^4\mu^2x^2d_2^3-4\alpha\beta^2\mu^2xd_1^2d_2-4\alpha\beta^2\mu^2xd_2^3+\alpha^2\mu^2d_1^2d_2+\alpha^2\mu^2d_2^3)e^{4\beta t}+(-8d_2\beta^4\mu^3x^2d_1^2\\
&&\hspace{0.6cm}-8d_2^3\beta^4\mu^3x^2-4id_2\beta^3\mu^3d_1^2-4id_2^3\beta^3\mu^3+8d_2\alpha\beta^2\mu^3xd_1^2+8d_2^3\alpha\beta^2\mu^3x-2d_2\alpha^2\mu^3d_1^2-2d_2^3\alpha^2\mu^3)e^{2\beta t}\\
&&\hspace{0.6cm}-(2\beta^2\mu^5d_1^4d_2+4\beta^2\mu^5d_1^2d_2^3+2\beta^2\mu^5d_2^5)e^{-2\beta t}+2d_2\mu\beta^4e^{6\beta t}+d_2\beta^4e^{8\beta t}+d_2^5\mu^4\beta^2+d_2\mu^4d_1^4\beta^2+2d_2^3\mu^4d_1^2\beta^2.
\end{eqnarray*}

From the above expressions (\ref{xt-3-24}) and (\ref{xt-3-25}), we can find that the first-order rogue wave of the coupled G-P equation has also a curved background, which are observed in other variable-coefficient models. Additionally, the parameter $\beta$ determines the degree of this curved background, that is, the background  will become steeper as $\beta$ increases; the background will become flatter as $\beta$ decreases. This phenomenon can be observed in Fig.8 and Fig.9. For the same reason that $\rho_j=[(\lambda-\lambda_j)(\lambda-\lambda_j^{*})^2]^{-\frac{1}{3}}$  occur in the above DT of the system (\ref{xt-3-1}), we can not give the determinant representations of DT and of its solutions $(q_1[N],q_2[N])$. Thus, we can not also utilize He's method \cite{3-43,3-44,3-45} to construct higher-order rogue waves of Eq.(\ref{xt-3-1}), meanwhile, the generalized Darboux transformation \cite{3-39} is not directly used to construct higher-order rogue waves. We haven't found appropriate method to give its higher-order rogue waves.

%
%

\section{Multi-component coupled  G-P equation and its Darboux transformation}
 In plasma, if the waves propagate in  more than two fields, we need to generalise the coupled G-P equation (\ref{xt-3-1}) to  multi-component \cite{3-25} to determine this case, which reads as
\begin{equation}
iq_{jt}+q_{jxx}+2\mu^2(\sum_{i=1}^{N}|q_i|^2)q_j+(i\beta-\alpha x+\beta^2 x^2)q_j=0~(j=1,2,3,\cdots,N),\label{xt-3-26}
\end{equation}
where $\alpha,~\mu,~\beta$ are all real constants, Eq.(\ref{xt-3-26}) admits the following Lax pair with $(N+1)\times(N+1)$ matrices
\begin{eqnarray}
&&\Psi_{x}=U\Psi=(i\lambda U_0+\mu U_1)\Psi,\label{xt-3-27}\\
&&\Psi_{t}=V\Psi=(2i\lambda^2 U_0+2\lambda( -i\beta x U_0+\mu U_1)+iV_1)\Psi,\label{xt-3-28}
\end{eqnarray}
where
\begin{gather*}
U_0=\begin{pmatrix}-1& & & & & & & & \\
 &1& & & & & & &\\
 & & &1& & && & & \\
 & & & &\ddots&& & & & \\
 && & & & & & &1& \\
& & & & & & & & &1\\
     \end{pmatrix}, \quad
U_1= \begin{pmatrix}0& Q_1&Q_2    &\cdots&Q_{N-1}&Q_N\\
                -Q_ 1&   0& &      &       &\\
                -Q_2&     &0     &      &       &\\
                \vdots&   &       &     \ddots &       &\\
           -Q_{N-1}^{*}&   &       &     &    0   &\\
           -Q_{N}^{*}  &   &       &     &       &0
\end{pmatrix},
\end{gather*}
\begin{gather*}
V_1=\begin{pmatrix}\mu^2\sum_{i=1}^{N}|Q_i|^2{-}\frac{\alpha x}{2} & \mu Q_{1x}{+}2i\mu\beta xQ_1 & \mu Q_{2x}{+}2i\mu\beta xQ_2 & \cdots & \mu Q_{(N{-}1)x}{+}2i\mu\beta xQ_{(N{-}1)} & \mu Q_{Nx}{+}2i\mu\beta xQ_N\\
                                      \mu Q_{1x}^{*}{-}2i\mu \beta xQ_1^{*} & {-}\mu^2|Q_1|^2{+}\frac{\alpha x}{2} & {-}\mu^2 Q_2Q_1^{*} &\cdots & {-}\mu^2Q_{(N{-}1)}Q_1^{*}&{-}\mu^2Q_{N}Q_1^{*}\\
                                       \mu Q_{2x}^{*}{-}2i\mu \beta xQ_2^{*} & {-}Q_1Q_2^{*}&{-}\mu^2|Q_2|^2{+}\frac{\alpha x}{2} &\cdots & {-}\mu^2Q_{(N{-}1)}Q_2^{*}&{-}\mu^2Q_{N}Q_2^{*}\\
                                       \vdots& \vdots & \vdots & \vdots &\vdots & \vdots \\
                                       \mu Q_{(N{-}1)x}^{*}{-}2i\mu \beta xQ_{(N{-}1)}^{*} & {-}\mu^2Q_1Q_{(N{-}1)}^{*} & {-}\mu^2Q_2Q_{(N{-}1)}^{*} & \cdots & {-}\mu^2|Q_{(N{-}1)}|^2{+}\frac{\alpha x}{2}& {-}\mu^2 Q_NQ_{(N{-}1)}^{*}\\
                                       \mu Q_{Nx}^{*}{-}2i\mu\beta xQ_N^{*} & {-}\mu^2Q_1Q_N^{*} & {-}\mu^2 Q_2Q_N^{*} & \cdots & {-}\mu^2Q_{(N{-}1)}Q_N^{*} & {-}\mu^2|Q_N|^2{+}\frac{\alpha x}{2}
\end{pmatrix},
\end{gather*}
with $\Psi=(\psi_1,\psi_2,\psi_3,\cdots,\psi_{(N+1)})^T$, $Q_j=q_je^{-\frac{i\beta x^2}{2}}~(j=1,2,3,\cdots,N)$.

Similarly, let $\Psi(\xi_1) =(\psi_1(\xi_1),\psi_2(\xi_1),\psi_3(\xi_1),\cdots,\psi_{(N+1)}(\xi_1))^T$ be a special vector solution of Lax pair (\ref{xt-3-27}) and (\ref{xt-3-28}), with choosing the seed solution of Eq.(\ref{xt-3-26}) $q_j= q_j[0]~(j=1,2,3,\cdots,N)$ at $\lambda=\lambda_1=\frac{\alpha}{4\beta}+\xi_1e^{-2\beta t}$. Utilizing the method of constructing Darboux transformation of the two-component coupled G-P equation, we can obtain the first-order DT of Eq.(\ref{xt-3-26})
\begin{eqnarray}
&&\widetilde{\Psi}=T\Psi,\quad T=\rho_1(\lambda)(\lambda I-H[1]\Lambda_1H[1]^{-1}),\label{xt-3-30}\\
&&\widetilde{q_j[1]}=q_j[0]+\dfrac{2i(\lambda_1^{*}-\lambda_1)\psi_1(\xi_1)\psi_{j+1}(\xi_1)^{*}}{\sum_{j=1}^{N+1}|\psi_j(\xi_1)|^2}e^{\dfrac{i\beta x^2}{2}},\label{xt-3-31}
\end{eqnarray}
where
\begin{eqnarray*}
&&\rho_1(\lambda)=[det(\lambda I-H[1]\Lambda_1H[1]^{-1})]^{-\frac{1}{N+1}}\\
&&\hspace{0.84cm}=[(\lambda-\lambda_1)(\lambda-\lambda_1^{*})^N]^{-\frac{1}{N+1}},\\
&&\Lambda_1= \begin{pmatrix}\lambda_1& & & & & & & & \\
 &\lambda_1^{*}& & & & & & &\\
 & & &\lambda_1^{*}& & && & & \\
 & & & &\ddots&& & & & \\
 && & & & & & &\lambda_1^{*}& \\
& & & & & & & & &\lambda_1^{*}\\
     \end{pmatrix},\\
&&H[1]=\begin{pmatrix}\psi_1(\xi_1) & \psi_2^{*}(\xi_1) &\psi_3^{*}(\xi_1) &\cdots &\psi_N^{*}(\xi_1)&\psi_{(N+1)}^{*}(\xi_1)\\
                       \psi_2(\xi_1) & -\psi_1^{*}(\xi_1)  & 0&\cdots&0&0\\
                       \psi_3(\xi_1)&0&-\psi_1^{*}(\xi_1)&\cdots&0&0\\
                       \vdots&  \vdots&\vdots&\vdots&\vdots&\vdots\\
                       \psi_{N}(\xi_1)&0&0&\cdots&-\psi_1^{*}(\xi_1)&0\\
                       \psi_{(N+1)}(\xi_1)&0&0&\cdots&0&-\psi_1^{*}(\xi_1) \end{pmatrix}.
\end{eqnarray*}
here, $I$ is the $(N+1)\times(N+1)$ identity matrix, besides, both $\Lambda_1$ and $H[1]$ all denote $(N+1)\times(N+1)$ matrices. Through the similar procedures in Eq.(\ref{xt-3-29}), the \emph{N}-order Darboux transformation of multi-component coupled G-P equation can also be given, and we omit it here. When $\emph{N}=1$ in Eqs.(\ref{xt-3-30}) and (\ref{xt-3-31}), we can get the  first-order Darboux transformation for one-component G-P equation here. 

\section{Conclusion}
In \cite{3-19} and \cite{3-24}, the Darboux transformation of one-component G-P equation was constructed, we verify this DT by Maple software and find that it is incorrect, because it does not hold $t$-part of Lax pair. Actually, the Lax pair of the G-P equation in Eq.(\ref{xt-3-26}) is nonisospectral with $\lambda_t\neq0$. In \cite{3-19} and \cite{3-24}, the authors constructed the Darboux transformation based on standard AKNS hierarchy with isospectral Lax pair, however, the DT of nonisospectral cases are essentially different from the isospectral ones.

In this paper, based on Zhou's method\cite{3-32,3-33}, we reconstruct the Darboux transformation of multi-component coupled G-P equation, especially, the two-component couple G-P equation is discussed in detail. In  the Lax pair (\ref{xt-3-2}) and (\ref{xt-3-3}), the spectral parameter $\lambda$ holds $\lambda=\frac{\alpha}{4\beta}+\xi e^{-2\beta t}$, here, $\xi$ is an arbitrary constant and we take it as the new spectral  parameter\cite{3-19}, so we give the infinitely-many conservation laws of Eq.(\ref{xt-3-1}) to make sure its integrability. Utilizing the Darboux transformation which is constructed by us, the nonautonomous solitons, breather and first-order rogue wave have been presented. In  nonautonomous solitons, the parameter $\mu$  affects the amplitude of soliton and $\beta$ changes the absolute value of propagation velocity $v_i$'s~$(i=1,2,3,\cdots,N)$. Both propagation direction of each soliton and the value of $v_i$'s are all determined by the real part of the new spectral parameters $\xi_i$'s~$(i=1,2,3,\cdots,N)$. These results are all discussed  in one nonautonomous soliton. For two nonautonomous solitons, choosing $Re(\xi_1)\neq Re(\xi_2)$ in Eqs.(\ref{xt-3-15}) and (\ref{xt-3-16}), the two interactional solitons are obtained in Fig.5, and if we take $Re(\xi_1)=Re(\xi_2)$, the two-soliton bound state is also exhibited in Fig.6. In order to get breather and rogue wave, the seed solutions of Eq.(\ref{xt-3-1}) should be selected as  nonzero background and this seed solutions are curved background. Based on this curved background, we get nonautonomous breather and first-order rogue wave. For this reason, the amplitude of the breather becomes small till being zero as $t$ increases. In first-order rogue wave, the parameter $\beta$ determines the degree of this curved background, which can be seen from Fig.8 and Fig.9. Our results further reveal the striking dynamic structures of analytical solutions in a nonautonomous coupled system, and we hope these results in this paper will be verified in physical experiments in the future. Due to  $\rho_j=[(\lambda-\lambda_j)(\lambda-\lambda_j^{*})^2]^{-\frac{1}{3}}~(j=1,2,3,\dots,N)$ exist in the j-step DT of Eq.(\ref{xt-3-1}), we can not give the determinant representations of DT and solutions. For the same reason, we can not directly construct higher-order rogue wave of Eq.(\ref{xt-3-1}) and we haven't found appropriate method to give its higher-order rogue wave. Besides, we will attempt to construct higher-order rogue wave of this coupled G-P equation in our future work.

\section*{Acknowledgment}
The project is supported by the Global Change Research Program of China(No.2015CB953904),
National Natural Science Foundation of China (No.11675054, 11435005). and Shanghai Collaborative
Innovation Center of Trustworthy Software for Internet of Things (No.ZF1213).

\end{document}